# The WebStand Project

*Benjamin Nguyen*
University of Versailles
benjamin.nguyen@prism.uvsq.fr

*François-Xavier Dudouet*
Université Paris-Dauphine
fxdudouet@dauphine.fr

*Dario Colazzo*
Université Paris-XI
dario.colazzo@lri.fr

*Antoine Vion*
University of Aix-Marseille II
antoine.vion@univmed.fr

*Ioana Manolescu*
INRIA-Futurs
loana.manolescu@inria.fr

*Pierre Senellart*
ENST
pierre@senellart.com

*"You affect the world by what you browse."*

*- Tim Berners-Lee*

## ABSTRACT

In this short paper we present the state of advancement of the French ANR *WebStand* project. The objective of this project is to construct a customizable XML based warehouse platform to acquire, transform, analyze, store, query and export data from the web, in particular mailing lists, with the final intension of using this data to perform sociological studies focused on social groups of World Wide Web. We are currently using this system to analyze the standardization process of the W3C, through its social network of standard setters.

## 1. INTRODUCTION

In this poster presentation, available in a longer version at [4] and [6], we describe our platform, *WebStand[1]*, currently under development, to be used by sociologists when studying information found on the Web, and in particular analyzing social behavior on mailing lists, forums or any place in which (tracked) discussions take place on the Web. Our current focus is the analysis of the W3C standardization mechanism.

Indeed, Information Technology is only just receiving attention from sociologists, and our goal is to create new tools for sociologists to assess and analyze this domain.

Our approach, when designing our initial platform architecture, was to consider, in conjunction with sociologists what sort of information they whished to obtain, and what sort of analysis they wanted to run. A preliminary study led us to the following conclusions:

Traditionally, sociological data consist of reports, questions and interviews. On the contrary, in the Web context, the data manipulated is **electronic**: mailing lists, homepages, and institution or company pages. Our goal is to discover, extract, and analyze actors of this field, their positions, their relationships, and their influence, etc. All this data is particularly adapted to automatic processing.

The *WebStand* approach is based on the use of a semi-structured temporal XML content warehouse to store the data, and graphically generated XQueries to analyze it. Let us stress that our warehouse aims to cover the whole Extract Transform and Load (ETL) scope of a sociological application. Our goal in this short paper is to focus on the architecture and temporal model of our application, briefly present the modules already developed, and give some sociological results that illustrate the sort of information that we can calculate easily.

## 2. INNOVATION AND SYSTEM STRONG POINTS

The general principle of an application focused integration platform is not new, in this section we describe the particularities of the data we manage, and the end-users, in order to stress the specific and novel characteristics of our system. These lie in several aspects:

*Native tree based data* (email lists or online forums)
XML and XQuery allow us to perform specific tree based query operations such as finding discussion patterns i.e., persons $P_1$ and $P_2$ are said to be *in discussion* if we find more than one pattern where messages posted by $P_1$ are fathers of messages posted by $P_2$ in a given thread, and vice-versa, or simply counting the depth of each discussion (transitive closure in relational). The use of XML is pivotal to our system, and we are currently developing XQuery optimization algorithms, tailored to our specific classes of queries.

*User friendly GUI, incorporating XML Schemas*
Although XQBE[1] implements support of XML Schemas, Schema based querying is not its main focus. In our case, experience with sociologists (therefore non expert users) has shown us that defining an initial *a priori* XML Schema to give some sort of shape to the results of the query. This a priori result schema is constructed using a user friendly GUI, and the query is then constructed by dragging elements of the data schema towards elements of the result schema. The real schema validating the result document is inferred from the data schema, the a priori result schema, and the choices made by the user when interacting with the GUI to construct his query, and of course, the query code is generated automatically.

---

[1] This work is partially funded by the French ANR-JCJC-05 "WebStand".

Query results can be physically materialized in the database, or can be reused as views, a process that sociologists are largely familiar with, by using Access. Let us stress that the use of views largely simplifies the constructing of complex queries, by breaking them into a succession of views. We are currently working on the optimization of these combined queries.

*A novel approach to temporality, based on sourcing*
Source based temporal information finds its roots in the work of Buneman [7] on provenance and Widom on lineage [8]. A novel *source based temporal model* is implemented in our platform. While we do not have the room to explain it in detail, we give a brief description in Section 4. Typical sociological data that need to be stored is the following: "The researcher Ann Onymous learns on 1-1-2008 that the French journal *Le Monde* published in its 4-3-2006 issue that John Doe joined XML Corp. on 10-10-2001." As shown in this example, traditional validity time and transaction time are not powerful enough to capture such information, that needs to be stored in conjunction with its origin. We refer to [6] for more details on this subject; which is still under progress.

## 3. ARCHITECTURE AND MODULES

The WebStand architecture is shown in Figure 1. WebStand is implemented in Java, and is running using the JDBC compatible MonetDB-XQuery [3] database. The modules developed include (a) a simple schema editor (b) an XML querying and visualization tool, geared towards mailing lists analysis, (c) a CV crawler and analyzer based on the Exalead crawler[2], (d) an email list crawler, extractor and cleaner, (e) a conversion module to export the data to external sociology applications. Current extensions of the system concern mainly improving the ergonomics of these modules and improve application tailored web data acquisition modules, that are currently rule based information extraction of pages retrieved by the exalead.com crawler.

Although we use MonetDB XQuery database to store the data, in some cases where the queries can not be run (such as queries using temporal functions) we use Saxon-B to compute the result.

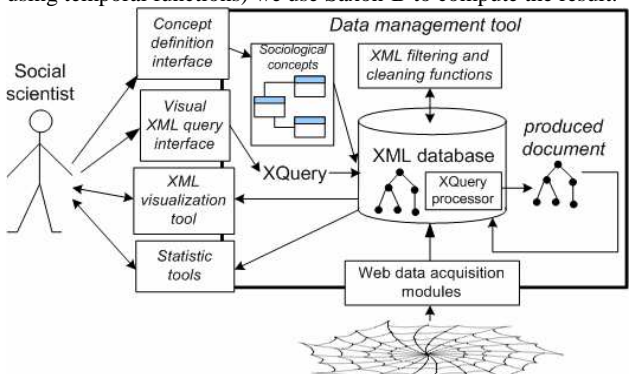

**Figure 1- WebStand Architecture**

The global use case is the following: a social scientist defines the concepts he is interested in, choosing from already existing concept (such as *person* or *email*) that can be extended with his own. This sociological model is (for the moment manually) translated into an XML Schema, used to store information extracted from the web by the acquisition modules. This XML Schema is also used to help the sociologist generate graphical queries, using a QBE-like interface, developed in our visualization and query tool. We used QBE rather than XQBE [1] due to the widespread use of Microsoft Access by the sociologists we work with, but we are considering alternative query interfaces based on XQBE. *WebStand* also provides simple XSL to export XML data in many formats used in the sociology world, although in a forseable future, we envision these applications to be all compatible with a simple XML format.

A preliminary study using our tool on 8 public mailing lists, related to XQuery and XML Schema has been performed. We are currently working on analyzing the data provided by all the public mailing lists of the W3C working groups.

## 4. SOCIOLOGICAL RESULTS

The corpus we focused on consist of 20.697 emails posted over the course of 4 years (from April 02 – to April 06) by about 3000 different "physical" people (i.e., after grouping emails together based on our heuristics, we identified 2923 different "entities"), analyzed according to activism on the lists and their participation in the writing of working drafts or recommendations. These emails originated from approximately 2000 different domains (Institutions or Internet Service Providers, our heuristics led us to 2076 different domains)

It is possible to run any query on this data, we show here simple aggregate results obtained to illustrate simple yet non the less valuable participation information.

Table 1 illustrates activism within W3C. It contains anonymized data showing the number of posts made by a single person: the top poster scored 1077 different posts. It is interesting to note that only 4 posters posted over 500 messages, and that only 500 posters out of 3000 posted over 5 messages. Turning to table 2, posts are now grouped by institution. We see that Microsoft and IBM dominate the playfield with Oracle tailing them. W3C posters are of course present. It is interesting to point out that the posts made by software AG all came from the same person, who went on to create his own XSL/XQuery company. Public research organizations such as universities are only represented by Edinburgh, UK, and although some public researchers post via their personal address (yahoo, aol, etc.) their participation is low, as show in Table 3, which illustrates the number of different posters, by domain name: 111 different people posting from yahoo.com posted 288 messages. On the other hand companies in terms of participation are once again IBM, Microsoft, and Oracle. We can see that Microsoft participant were extra-active, since only 20 people (compared to IBM's 35) posted nearly twice the number of emails. On the other end of the scope, universities and public research institutions are unable to mobilize a large number of active participants.

| #   | Posts |
|-----|-------|
| 1.  | 1077  |
| 2.  | 730   |
| 3.  | 683   |
| 4.  | 604   |
| 5.  | 423   |
| 6.  | 385   |
| 7.  | 373   |
| 8.  | 318   |
| 9.  | 225   |
| 10. | 223   |
| 11. | 207   |
| 12. | 203   |
| 13. | 198   |
| 14. | 197   |
| 15. | 169   |

**Table 1- Post count per person**

| Institution | Posts |
|---|---|
| microsoft.com | 1547 |
| ibm.com | 978 |
| softwareag.com | 681 |
| w3.org | 623 |
| oracle.com | 564 |
| cogsci.ed.ac.uk | 555 |
| acm.org | 485 |
| mhk.me.uk | 425 |
| nag.co.uk | 318 |
| yahoo.com | 288 |
| aol.com | 259 |
| datadirect.com | 212 |
| sun.com | 206 |
| arbortext.com | 203 |
| metalab.unc.edu | 196 |
| CraneSoftwrights.com | 180 |
| hotmail.com | 168 |
| kp.org | 165 |
| jclark.com | 141 |
| bea.com | 125 |

**Table 2- Post count per institution**

Table 4 shows the number of technical reports signed by members of institutions that signed at least one XQuery related[2] recommendation. Once again we see that IBM outnumbers Microsoft by 2:1 both on the number of different authors and on the number of recommendations. Universities are also nearly non-existent. From a "neutral" sociologist point of view, these results point to the conclusion that corporations seem to dominate XQuery standard setting.

| Domain | Posters |
|---|---|
| yahoo.com | 111 |
| hotmail.com | 101 |
| w3.org | 99 |
| ibm.com | 35 |
| fake.invalid | 32 |
| excite.com | 27 |
| aol.com | 24 |
| microsoft.com | 20 |
| oracle.com | 20 |
| gmail.com | 18 |

**Table 3- Posters per domain**

---

[2] We selected 28 technical reports in the recommendation process that appeared in the discussions on the list.

| INSTITUTION | TYPE | # INDIV | REC. | W3C WG NOTES | DRAFTS |
|---|---|---|---|---|---|
| **IBM** | Corp | 11 | **8** | 2 | 3 |
| **Oracle** | Corp | 8 | **6** | 1 | 6 |
| **AT&T** | Corp | 2 | **4** |  | 3 |
| **Microsoft** | Corp | 5 | **4** |  | 2 |
| **Unknown** | n.a. | 2 | **3** |  |  |
| **Sun Microsystems** | Corp | 1 | **3** |  |  |
| **Data Direct Technologies** | Corp | 1 | **2** | 2 | 2 |
| **University of Edimbourg** | Uni | 2 | **2** | 1 |  |
| **Saxonica** | Corp | 1 | **2** |  |  |
| **Infonyte GmbH** | Corp | 1 | **1** | 2 |  |
| **Brown University** | Uni | 1 | **1** |  |  |
| **CommerceOne** | Corp | 1 | **1** |  |  |
| **Inso** | Corp | 1 | **1** |  |  |
| **Kaiser Permanente** | Org | 1 | **1** |  |  |
| **SIAC** | Corp | 1 | **1** |  |  |

**Table 4- Recommendation information**

Information used to create Table 4 was entered by hand using our temporal model detailed in section 4. We are currently in the process of automating authoring information from the versions (from WD to REC) of one W3C technical report found on the Web.

Other results that are produced by our system are *social graphs*, that indicate common participation on a thread, *answering profiles* that indicate with which other list participants a given person privileges discussion, we can not provide them here due to the fact these graphs are place consuming, but we give one example in the Appendix, and we also refer to [4] for more examples of these graphs.

## 5. EFFECTIVENESS AND SCALABILITY

We believe that once queries are generated automatically, their optimization is an orthogonal issue that can be dealt with by specific algorithms, and to this end we are currently working on the optimization and benchmarking of particular classes of queries. However, our prototype aims to show the feasibility of our approach. Tests so far have been applied to *real* collections of approximately 2GB of W3C emails, downloaded, parsed and stored in XML. Document sizes range from several KB to 100MB.

## 6. CONCLUSION

In this short paper, we present a brief overview of the architecture and functionalities of the *WebStand* platform and give some brief results of a study of the W3C. For more details on the sociological results, we refer to [4]. Our current experience shows that use of XML and XQuery through simple graphical interfaces simplifies the accessibility of XQuery to novice users, such as sociologists. We have not discussed here our temporal model, which is still under development but has allowed us to capture all the data collection situations that we have encountered so far, and it is our belief that such software can be used in various other sociological applications to analyse behaviors.

4. **APPENDIX**

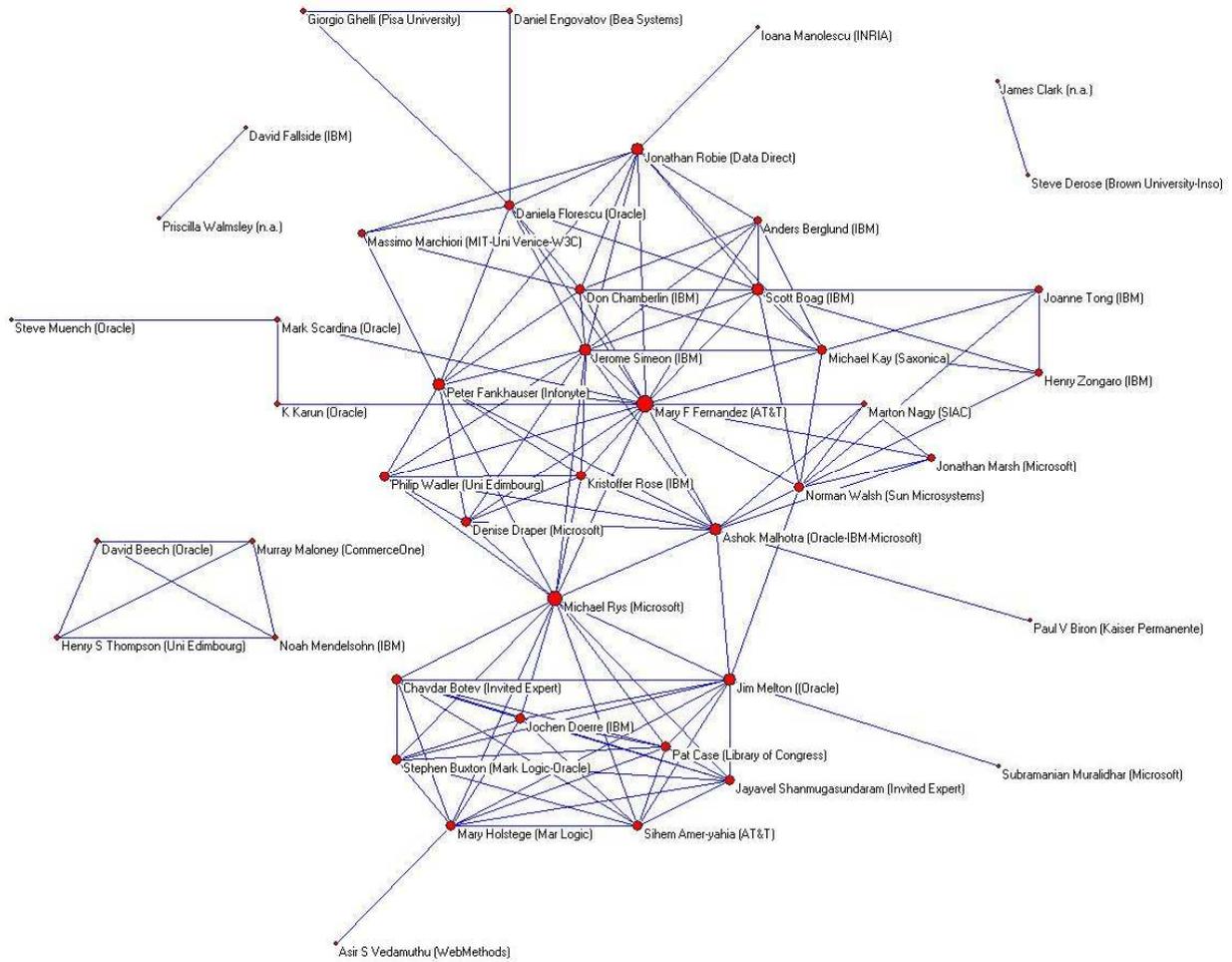

**Figure 2- Recommendations co-authors institutional mapping[3]**

---

[3] Memberships are established automatically and the data only takes into account the ones declared by authors on the formal recommendations they sign.